\documentclass[useAMS,usenatbib]{mn2e}
\usepackage{graphicx}
\usepackage{epsfig}

\title[Spinar Paradigm and Gamma Ray Bursts Central Engine]{Spinar Paradigm and Gamma Ray Bursts Central Engine}
\author[V.Lipunov and E.Gorbovskoy]{V.M. Lipunov$^{1,2,3}$ and E.S. Gorbovskoy$^{1,2,3}$\\
$^{1}$Sternberg Astronomical Institute, Moscow, Universitetsky pr. 13, Moskow 119992, Russia\\
$^{2}$Moscow State University, Moscow, Universitetsky pr. 13, Moskow 119992, Russia.\\
$^{3}$Moscow Union ``Optic'', Moscow, Valilov str 5/3, Moskow 119334, Russia.}
\begin{document}

\date{Accepted 5 September 15. Received 2007 August 31; in original form 2007 June 21}

\maketitle

\label{firstpage}

\begin{abstract}

A spinar is a quasi-equilibrium collapsing object whose equilibrium 
is maintained by the balance of centrifugal and gravitational forces and 
whose evolution is determined by its magnetic field. The spinar quasi 
equilibrium model was recently discussed in the context of extralong X-ray 
plateu in GRB (Lipunov {\&} Gorbovskoy, 2007).

We propose a simple non stationary three-parameter collapse model 
with the determining role of rotation and magnetic field in this paper. The 
input parameters of the theory are the mass, angular momentum, and magnetic 
field of the collapsar. The model includes approximate description of the 
following effects: centrifugal force, relativistic effects of the Kerr 
metrics, pressure of nuclear matter, dissipation of angular momentum due to 
magnetic field, decrease of the dipole magnetic moment due to compression 
and general-relativity effects (the black hole has no hare), neutrino 
cooling, time dilatation, and gravitational redshift. 

The model describes the temporal behavior of the central engine and 
demonstrates the qualitative variety of the types of such behavior in 
nature.

We apply our approach to explain the observed features of gamma-ray 
bursts of all types. In particular, the model allows the phenomena of 
precursors, x-ray and optical flares, and the appearance of a plateau on 
time scales of several thousand seconds to be unified. 

\end{abstract}

\begin{keywords}

black hole physics --- Physical Data and Processes, gravitation ---
Physical Data and Processes, magnetic fields --- Physical Data and
Processes, relativity --- Physical Data and Processes, gamma-rays:
bursts --- Sources as a function of wavelength, gamma-rays:
theory --- Sources as a function of wavelength
\end{keywords}

\section{Introduction}.

The interest toward magneto-rotational collapse has increased appreciably in 
recent years in connection with the gamma-ray burst problem. It is now 
believed to be highly likely that long gamma-ray bursts may be associated 
with the collapse of a rapidly rotating core of a massive star and short 
gamma-ray burst are most likely to be results of the coalescence of neutron 
stars, which can be viewed as the collapse of a rapidly rotating object. We 
already pointed out in our earlier papers (Lipunova, 1997, Lipunova {\&} 
Lipunov, 1998) the likely multivariate nature of, e.g., the coalescence of 
two neutron stars or neutron stars and black holes (``mergingology''), which 
may give rise to various forms of the temporal behavior of gamma-ray bursts. 
This is possibly corroborated by the recent complex classification of 
gamma-ray bursts (Gehrels et al., 2006).

Moreover, observations of the so-called precursors and x-ray flare certainly 
point to the complex nature of the operation if their central engines 
(Lazzati, 2005; Chincarini et al., 2007). ROTSE (Quimby et al., 1996a,) and 
MASTER (Lipunov et al., 2007) facilities observed optical flares in a number 
of cases.

All this triggers (mostly numerical) theoretical studies of collapse with 
the dominating role of rotation. Numerous attempts have been undertaken in 
order to incorporate effects due to rotation and magnetic fields in 
numerical computations, which are very difficult to understand intuitively 
and at the same time are extremely approximate because of the complex nature 
of the problem (Gehrels et al., 2006, Moiseenko et al., 2006; Duez et al., 
2005, 2006).

Recently, (Lipunov {\&} Gorbovskoy 2007) showed that spinar paradigm 
naturally explains not only the phenomenon of early precursors and flares, 
but even extraordinarily long x-ray plateaux. 

In this paper we propose a pseudo-Newtonian theory of collapse based on a 
simple analytical model, which allows the maximum number of physical effects 
to be incorporated. 

We use our model to interpret the data of observations of precursors 
(Lazzati, 2005), X-ray flares (Chincarini ey al., 2007), and some 
interesting gamma-ray bursts.

\begin{figure}[ht]
\psfig{figure=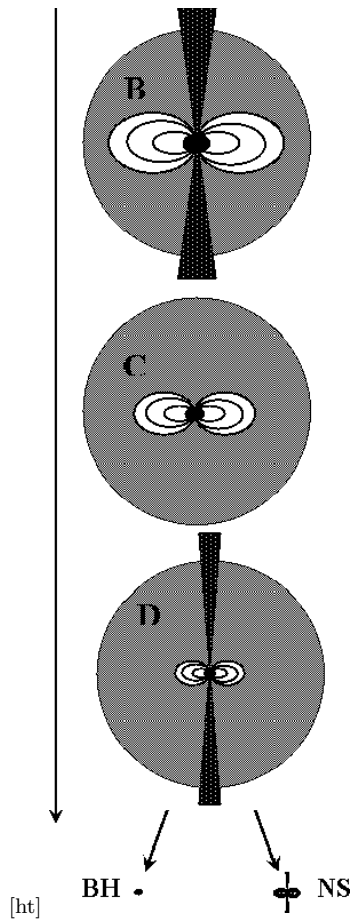,width=40mm}
\caption{
 Schematic view of the collapse of the rapidly rotating magnetized core of 
a massive star. Gray and black shaded areas show the envelope and core of the star, respectively. 
Before the collapse the size of the star is on the order of several solar radii and its iron core 
is one hundred times smaller (stage A). During the collapse centrifugal forces increase most rapidly,
resulting in the formation of a spinar (stage B). Its formation is accompanied by anisotropic release 
of energy. Because of the dissipation of angular momentum the spinar decelerates and contracts (stage C). 
Its luminosity increases and a new jet forms whose energy release reaches its maximum near the gravitational 
radius. Depending on the core mass, the process results in the formation of a neutron star or an extremely 
rotating black hole.}
\end{figure}

\section {The Spinar Model.}

The importance of incorporating magneto-rotational effects in collapse 
models was first pointed out in connection with the problem of quasar energy 
release and evolution (Hoyle and Fowler, 1963; Ozernoy, 1966; Morison, 1969; 
Ozernoy and Usov, 1973), and that of the ejection of supernova shells 
(Bisnovaty-Kogan; 1971, LeBlance {\&} Wilson 1970). 

In particular, it was pointed out that the collapse of a star having 
substantial angular momentum may be accompanied by the formation of a 
quasi-static object -- a spinar -- whose equilibrium is maintained by 
centrifugal forces. Ostriker (1970) and Lipunov (1983) assumed the existence 
of low-mass spinars with close-to-solar masses. Lipunov (1987) made a 
detailed analysis the spin-up and spin-down of spinars in the process of 
accretion.

Lipunova (1997) developed a spinar model incorporating relativistic effects 
(which include the disappearance of magnetic field during the formation of a 
black hole), gave an extensive review of the research on the spinar theory, 
and tried to apply the spinar model to the gamma-ray event. 

A spinar can be viewed as an intermediate state of a collapsing object whose 
lifetime is determined by the time scale of dissipation of the angular 
momentum. As Lipunova {\&} Lipunov (1998) pointed out, the centrifugal 
barrier could explain the long (from several seconds to several hours) 
duration of the process of energy release in the central engines of 
gamma-ray bursts. It is remarkable that as it loses angular momentum a 
spinar (unlike, e.g., a radio pulsar) does not spin-down, but, on the 
contrary, spins up and this effect results in the increase of luminosity, 
which is followed by the luminosity decrease because of the disappearance of 
magnetic field, relativistic effect of time dilatation, and gravitational 
redshift near the event horizon.

Lipunova (1997) analyzes a model of a spinar in vacuum, which is justified 
for two neutron stars. However, in the case of a collapse of a core of a 
massive star the spinar is surrounded by the star's envelope and matter 
outflowing from its equator. We analyzed the interaction of a spinar with 
the ambient plasma in our earlier paper Lipunov (1987), from where we adopt 
the law to describe the dissipation of the spinar angular momentum .

Recently, Lipunov {\&} Gorbovskoy (2007) developed a stationary spinar 
model, which allows for relativistic effects and maximum possible 
dissipation of the angular momentum of the spinar.

Below we abandon the quasi-stationary analysis and construct a 
non-stationary model of rotational collapse.

\section {Spinar scenario of magneto-rotational collapse. Collapse of a rapidly rotating core.}

Let us now qualitatively analyze the magneto-rotational collapse of a 
stellar core of mass $M_{core}$ and effective Kerr parameter (Thorne et al., 
1986)
\begin{equation}
a_0 \equiv \frac{I\omega _0 c}{GM_{core}^2 }
\end{equation}

(here $I = k M_{core} R_{0}^{2}$ is the moment of inertia of the core; $\omega$ 
is the angular velocity of rotation, and $c$ and $G$ are the speed of light and 
gravitational constant, respectively), and magnetic energy $U_{m}$.

In the case of conservation of the core angular momentum (which, of course, 
will be violated in our scenario), $a$ remains constant.

Let $\alpha_m$ be the ratio of the magnetic energy of the core to its 
gravitational energy:

\begin{equation}
\alpha _m \equiv \frac{U_m } {{GM_{core}^2 } / R_A}
\end{equation}

The total magnetic energy can be written in terms of the average magnetic 
field $B$ penetrating the spinar:

\begin{equation}
U_m =\frac{B^2}{8\pi }\frac{4}{3}\pi R^3=\left(\frac{1}{6}\right)B^{2}R^{3}
\end{equation}

Note that in the approximation of magnetic flux conservation ($ÂR^{2} = const$), the 
magnetic-to-gravitational energy ratio remains constant during the collapse: 
$\alpha _m=const, U_m \propto R^{-1}$ without considering 
general-relativity effects.

Let the initial Kerr parameter $a_{0} > 1$. In this case, direct formation of a 
black hole is impossible and the process of collapse breaks into several 
important stages (see Fig.1.):

\paragraph *{A). Loss of stability by the core and free fall}

The time scale of this stage is on the order of the free-fall time

\begin{equation}
T_A =\sqrt {\frac{R_A ^3}{GM_{core} }} \sim 100s
\end{equation}

where $R_{A}$ is the initial radius of the stellar core. Energy is virtually 
not radiated during the collapse, and gravitational energy transforms into 
kinetic, rotational, and magnetic energy of the core.

\begin{figure}[ht]
\psfig{figure=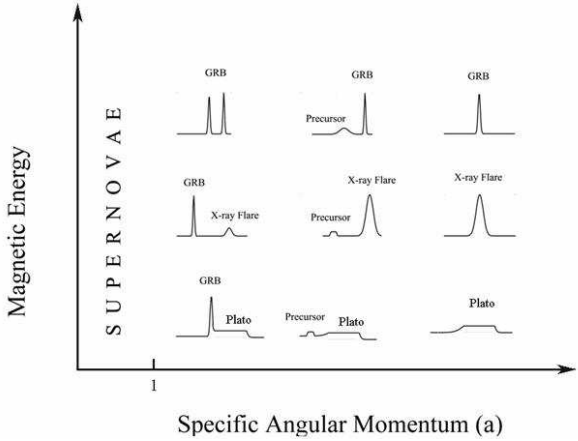,width=80mm}
\caption{
 Qualitative variation of the characteristics of a gamma-ray 
burst and the accompanying phenomena shown on the magnetic field --- effective Kerr 
parameter diagram.}
\end{figure}

\paragraph *{B). Halt of the collapse by centrifugal forces.}

Centrifugal forces stop free-fall collapse at the distance where 

\begin{equation}
\omega^2R_B ={\frac {GM_{core}}{{R_B^2}}}
\end{equation}

It follows from this that the initial spinar radius is approximately equal 
to:

\begin{equation}
R_{B}=a^{2}GM_{core}/c^{2}=a^{2}{Rg/2}
\end{equation}

In this process, half of the gravitational energy is released:

\begin{equation}
E_B=\frac{GM^2}{2R_B}-\frac{GM^2}{R_B} \approx \frac{GM^2}{2R_B} = \frac{1}{2a^2}M_{core}\,c^2
\end{equation}

if the energy is sufficient to ``penetrate'' the stellar envelope, i.e., if 
the momentum imparted to a part of the shell exceeds the momentum 
corresponding to escape velocity. Let a part of the energy be converted into 
the energy of the jet $(\beta E_j = \beta E_B)$

\begin{equation}
\frac{\beta E_B}{v_j} > \theta_B^2 M_{shell} \sqrt {\frac{2GM}{R_{shell}}}
\end{equation}

In this case a burst of hard radiation occurs. 

We now substitute the burst energy (formula (7)) and spinar radius (6) into 
condition (8) to derive the ``penetration'' condition for the first jet:

\begin{equation}
1<a_0 <\frac{1}{\theta _B ^2}\frac{M_{core} }{M_{Shell} }\frac{C}{V_p } 
\end{equation}

where $V_{p}$ is the escape velocity at the surface of the stellar envelope. 
In real situations $V_{p}~=~2000~-~3000~km/s, \frac{M_{core}}{M_{Shell}} \sim {\frac{1}{10}}-{\frac{1}{3}}$
, and almost 
everything is determined by the jet opening angle. This simple estimate 
shows that the first penetration is highly likely even in the case of a 
large jet opening angle.

Because of the axial symmetry, the burst must be directed along the rotation 
axis and have an opening angle of $\theta _B ^2$. The duration of this stage 
is determined by the time it takes the jet to emerge onto the surface 
($R_{shell} \sim 10-30s$) and the character of cooling governed by the structure of the 
primary jet and envelope.

The gamma factor of a jet emerging at the surface of the star can be 
approximately estimated using the energy conservation law (see a review by 
Granot, 2007):
\begin{equation}
E_{jet} \approx \Gamma ^2\theta _B^2 M_{shell} 
c^2 \textnormal {  and  } \Gamma \sim E_{50}^{1/2} 
\theta _B^{-1} \left(\frac{M_{shell} }{M_\odot}\right )^{-1/2}
\end{equation}

Here $E_{50}=E_{B}/10^{50}$ erg/s -- jet energy. 

The character of the spectrum is determined by the gamma factor of the jet. 

If the initial Kerr parameter is large ($a\gg1$) then energy 
$E_{B} \ll M_{core}c^{2}$ and the emerging jet is nonrelativistic allowing the 
event in question to be viewed as a precursor like it was done by 
Ramirez-Ruiz et al. (2002) and Wang {\&} Meszaros (2007). Its spectrum can 
be estimated by the blackbody formula (eq. 16 in Wang {\&} Meszaros, 2007):
$$
T\sim 15L_{50}^{1/8} R_{11}^{-1/4} Kev
$$
If the initial Kerr parameter is close to unity then the energy of the burst 
is high and the jet acquires a high gamma factor after penetration so that 
the flare should be interpreted as a gamma-ray burst. Although the jet that 
penetrates the star may be subrelativistic, however, a higher gamma-ray 
factor jet is to flood the already formed channel (the central engine 
continues to operate!). It is this evolved jet that should produce the 
gamma-ray burst provided that the spinar size is close to the gravitational 
radius.

We do not discuss the parameters of the jet, because this issue been 
addressed repeatedly by different authors ( see reviews by Granot (2007) and 
Piran (2005)). 

Only future numerical computations will make it possible to accurately 
determine the degree of anisotropy, i.e., e.g., the jet $\theta_B^2$. 

However, here we try to estimate the degree of anisotropy by determining the 
fraction of the spinar surface occupied by open field lines. Let us assume 
for a moment that the spinar has a dipole moment equal to $\mu$. Let us determine 
the Alfven radius $R_{Alfven}$ of the jet from the condition of the balance 
of the jet ram pressure and magnetic-field pressure:
\begin{equation}
\frac{L_{B}}{ (\theta _B^2R^{2}c)} \sim  
\frac{\mu^{2}}{8\pi\,R^{6}}
\end{equation}

We use this formula to derive the Alfven radius:
$$
R_{Alfven} \sim \theta _B ^{1/2}(c\mu^2/2L_b )^{1/4}
$$ 
We further assume that $\mu=BR^{3}/2$, 
\textit{where B is the intensity of magnetic field at the pole of the spinar,
 to obtain}

$$
R_{Alfven} \sim 3 \times 10^{8}{cm} (\theta _B /0.01)^{1/2}B_{15}^{1/2} R_7^{1/2}\,
 R_7 L_{50}^{-1/4} \gg R_7 
$$

Here $B_{15}=B/10^{15} Gs, R_{7} = R_{B}/10^{7 }cm.$

It is evident that all field lines passing inside this radius are closed. We 
use the approximation of the dipole field line equation to determine the 
size of polar regions enclosing open field lines:
$$
\theta_{polar}\approx (R_B /R_{Alfven} )^{1/2}\approx 0.03
(\theta _B /0.01)^{1/4}B_{15}^{1/4} R_7^{1/4} L_{50}^{-1/8} \ll 1
$$
Thus only $0.1{\%}$ of the spinar surface participate in the formation of the 
jet, implying a very high degree of anisotropy of the process considered.

The newly formed spinar then evolves until its collapse without losing its 
axial symmetry.

\paragraph *{C). Dissipative evolution of the spinar}

The spinar contracts as its angular momentum is carried away. Note that this 
process is accompanied by the increase of the velocity of rotation and 
luminosity of the spinar. At the same time, the magnetic dipole moment 
decreases and the luminosity stops increasing and begins decreasing. The 
energy release curve acquires the features of a burst.

The duration of this stage is determined by the moment of forces that carry 
away the angular momentum of the collapsar. In real situations turbulent 
viscosity and magnetic fields may play important part in the process. 

The corresponding dissipation time scale (the spinar life time) is:
\begin{equation}
t_C ={I_B \omega } / K_{sd}  
\end{equation}

where $K_{sd}$ is the characteristic torque of dissipative forces. It is 
clear that under the most general assumptions about the character of 
magnetic field the spin-down torque must be proportional to the magnetic 
energy of the spinar:
\begin{equation}
K_{sd} =\kappa_t U_m 
\end{equation}
where $\kappa_t$ is the dimensionless factor that determines how twisted 
magnetic field lines are via which the angular momentum is dissipated.

Correspondingly, the total time scale of the dissipation of angular momentum 
(spinar lifetime (9)) is equal to:
\begin{equation}
t_C \sim \frac{I\omega }{U_m }\sim \frac{GM_{core} a_0 ^3}{c^3\alpha _m 
\kappa _t }
\end{equation}

\paragraph *{D). Second burst}

Energy is released during dissipation, and the rate of this process 
increases progressively until general relativity effects --- redshift and 
disappearance of magnetic field come into play. 

As the luminosity increases, at a certain time instant the conditions of 
shell penetration (similar to condition (8)) become satisfied:
\begin{equation}
\frac{E_D}{c}>\theta _D ^2M_{shell} \sqrt {2\frac{GM}{R_{shell}}}
\end{equation}
A second jet appears whose intensity reaches its maximum near the 
gravitational radius. Note that the effective Kerr parameter tends to its 
limiting value for the extremely rotating Kerr black~hole:~$a\to{1}$.

The maximum luminosity can be written in terms of the dissipation of 
rotational energy near the gravitational radius:
\begin{equation}
L_D =\frac{M\,Rg^2\,\omega }{\alpha\, M\,c^2}\sim \frac{\alpha _m\, c^5}{G}
\end{equation}
It is better to write the condition of the penetration for the second jet in 
terms of pressure inequality:
\begin{equation}
\frac{L_D}{\theta_D^2\,cR^2}>\frac{GM^2}{R^4}
\end{equation}
Note that $\frac{c^5}{G}=10^{59}erg/s$ is the so-called natural luminosity.

Of course, formula (15) does not include gravitational redshift, decay of 
magnetic field, etc.

The time scale near the maximum is:
\begin{equation}
T_D \sim \frac{M_{core}\,Rg^2\omega}{U_m}=\frac {G\,M\,a^3}{c^3\alpha _m}
\end{equation}

Further fate of the star depends on its mass. If the mass exceeds the 
Oppenheimer--Volkoff limit the star collapses into a black hole. Otherwise 
(Lipunova {\&} Lipunov, 1998) a neutron star forms, which cools after 10 
seconds, continues to spin down in accordance with the following formula 

\begin{equation}
K={\mu^2}/R_l^3
\end{equation}

where $\mu$ is the magnetic dipole moment and $R_l =c/\omega$
is the radius of the light cylinder, and radiates as a common pulsar. In the case of constant magnetic field the 
luminosity of the pulsar should decrease in accordance with the following 
law:

\begin{equation}
L=\frac {\mu ^2\omega }{R_l ^3} \sim t^{-2}
\end{equation}

In the case of a coalescence of two neutron stars or a neutron star and a 
black hole the first stage (stage \textbf{A) }is very short, because the 
``fall'' begins at a distance of several gravitational radii. Because of 
gravity-wave losses the components of the binary first approach each other 
to the radius of the last stable orbit and then merge to form a spinar. A 
small burst may occur at the time of stellar merging immediately before the 
spinar forms. This burst has the energy of:
\begin{equation}
\Delta E=\frac{G\,(M_1 +M_2)^2}{R_B}-\frac{GM_1^2}{R_1}-\frac{GM_2^2}{R_2} 
\sim 0.1(M_1 +M_2)c^2
\end{equation}
The qualitative picture of magneto-rotational collapse considered here can 
be illustrated by the following scheme (see Fig. 2.) in the coordinates 
$U_{m}$ and $a$ --- the effective Kerr parameter.

The proposed scenario allows easy interpretation of the precursors and 
flares. In the case of large angular momentum $(a\gg1)$ the initial radius is large 
and, correspondingly, the energy release rate is low, allowing stage 
\textbf{B }to be interpreted as a precursor.

In the case of low angular momentum ($a >\sim 1$) the initial spinar radius is close to 
several gravitational radii and stage \textbf{B} must be interpreted as a 
gamma-ray burst, whereas the subsequent spinar burst \textbf{D }must be 
interpreted as a flare event.

It is remarkable that the time interval between the two bursts is always 
determined by the duration of dissipation of angular momentum (14), and, 
consequently, a rest-time measurement immediately yields a relation between 
the Kerr parameter and the fraction of magnetic energy:
\begin{equation}
\frac{\alpha _m }{a_0 ^5}=\frac{\Delta tc^5}{GM_{core} \kappa _t }\cong 
10^{-6}\frac{M}{10M_\odot }\Delta t_2 ^{-1}\kappa _t 
\end{equation}
where $\Delta t_2 =\Delta t/100s$.

Correspondingly, the characteristic magnetic field at the collapse time 
(near $Rg$) is equal to:

\begin{equation}
B=\left( {\frac{Rg}{R_{core} }} \right)^{-2}\sqrt {\frac{\alpha _m 
GM^2}{6R_{core}^4 }} \approx 2\cdot 10^{15}Gs\cdot \alpha _{-6}^{1/2}
{\left(\frac{M_{core} }{M_{\odot}}\right)}^{-3/2}
\end{equation}
where $\alpha _{-6} =\alpha _m /10^{-6}$.

The proposed scenario allows the observed variety of gamma-ray bursts, 
precursors, and flares to be reduced to just two parameters: magnetic field 
and initial angular momentum. 

Let us considered firstly two upper line of the diagram (Fig.2). In the case 
of weak magnetic field and large angular momentum (the right side of the 
middle line) the first burst is weak (because of the high centrifugal 
barrier) and the resulting jet does not penetrate the stellar envelope -- 
there are no precursors to be observed. This is followed by slow collapse 
(magnetic field is weak), which results in a weak x-ray rich burst. As the 
initial angular momentum decreases (we move leftward in the diagram along 
middle line) the energy released at the centrifugal barrier increases and 
the jet becomes capable of ``penetrating'' the stellar envelope. The first 
burst should act as a precursor. The precursor should be separated from the 
gamma-ray burst, because the time scale of the dissipation of angular 
momentum is long in the case of a weak field. As angular momentum decreases 
(we move further leftward along the horizontal middle line) the precursor 
energy increases and at {$a>\sim 1$} the precursor energy exceeds 
$10^{51-52} erg$ and it 
shows up as a gamma-ray burst with the subsequent collapse of the spinar 
leading to X-Ray flare or an X-ray plateau event (the bottom-line Lipunov 
{\&} Gorbovskoy, 2007) with more weak field. 

In the case of even stronger magnetic field, the flare approaches a 
gamma-ray burst, its energy grows and the flare itself becomes a part of the 
gamma-ray burst (the top-left corner). If we move rightward, angular 
momentum grows and the first flare loses energy and becomes a precursor 
close to the second flare, which, in turn, actually becomes a gamma-ray 
burst.

In the case of very large angular momentum (the top-right corner) the energy 
of the precursor is insufficient for penetrating the envelope and we have a 
burst without satellites. The duration of energy release increases with 
decreasing strength of magnetic field and the burst becomes softer (we come 
to the bottom-right corner) and turn into isolated long X-ray plateau. 

\begin{figure}[ht]
\psfig{figure=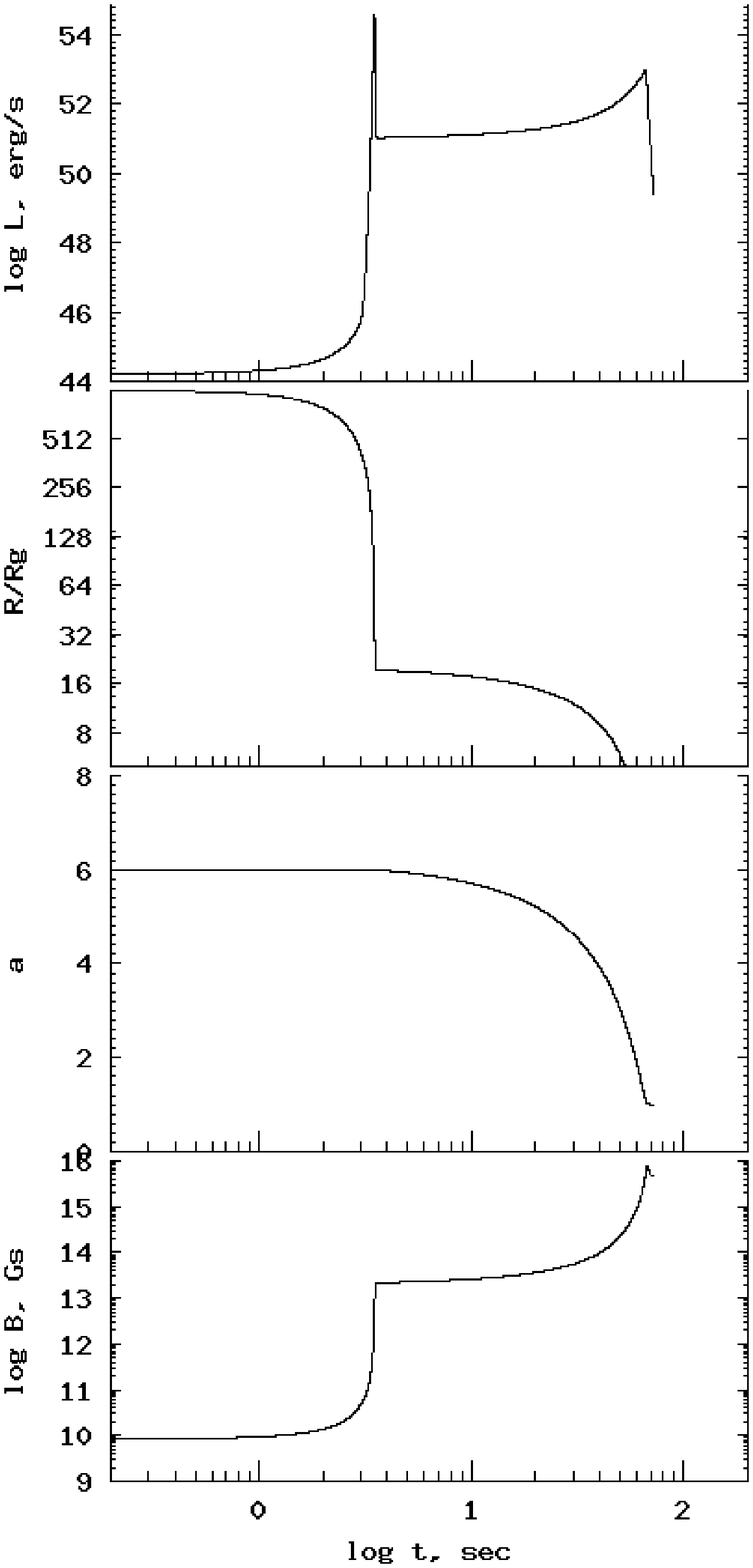,width=80mm,height=170mm}
\caption{
 Computation of the collapse of a 7 solar mass core with
effective Kerr parameter a$_{0}$=6 and magnetic-to-gravitational 
energy ratio $\alpha _{m}$=10$^{-4}$. From top
to down: energy release as viewed by an infinitely distant observer, radius,
effective Kerr parameter, and the average magnetic field strength.}
\end{figure}

\section {One point pseudo-Newtonian nonstationary Spinar Model of the 
magneto-rotational collapse.}

The aim of our model is to provide a correct qualitative and approximate 
description of magneto-rotational collapse, which would allow us to follow 
the evolution of the rate of energy release of the collapsing object and 
demonstrate the diverse nature of the central engine. Note that the spinar 
is born and dies in a natural way as a result of the solution of 
nonstationary problem.

Let us assume that at the initial time instant we have a rotating object (it 
may be a core of a massive star that has become unstable, or a merged 
neutron star, or the massive disk around a black hole). The object has the 
mass of $Ì$, radius $R_{core}$, angular momentum $I\omega$, dipole momentum 
$\mu _0$, and Kerr parameter $a_0$.

\subsection * {a). Dynamic Equation }

We write the equation of motion in the post-Newtonian approximation:
\begin{equation}
\frac{d^2R}{dt^2}=F_{gr} +F_c +F_{nuclear} +F_{diss}
\end{equation}
where $F_{gr}$ is the gravitational acceleration, $F_{c}$, the centrifugal 
acceleration, and $F_{nuclear}$, the pressure of matter.

Several attempts have been made to propose a pseudo-Newtonian potential to 
simulate the Kerr metrics (see Artemova et al., 1996 ). In our model we use 
effective acceleration in the form proposed by Mukhopadhyay (2002) for 
particles moving in the equatorial rotation plane:

\begin{equation}
F_{gr} =-\frac{GM}{x^3}\frac{(x^2-2ax+a^2)^2}{(\sqrt x (x-2)+a)^2}
\end{equation}

where $x=2R/R_g$. This formula corresponds to the potential of Paczynski {\&} 
Wiita (1980) for a nonrotating black hole.

Next terms:

\begin{equation}
F_c =\omega ^2R
\end{equation}

\begin{equation}
F_{nuclear} =\frac{1}{\rho }\frac{dP}{dr} \approx \frac{P}{\rho R}
\end{equation}

Pressure of gas, which includes thermal pressure, can be written as kinetic 
energy of particles computed using relativistic invariant (Zel'dovich, 
Blinnikov, Shakura 1980):

\begin{equation}
P\approx \rho (\sqrt {c^4+b\rho ^{2/3}+(Q/M)^2} -c^2)
\end{equation}

The second and third terms under the radical sign allow for the pressure of 
degenerate gas and thermal energy, respectively.

Let us now rename constant $b$: 
\begin{equation}
b=\left({\frac{4\pi }{3}}\right)^{2/3}\,G^2M_{Class}^{4/3}
\end{equation}
We actually use the formula for the pressure of partially degenerate Fermi 
gas with the contribution of thermal pressure. It is clear that the equation 
of real nuclear matter cannot be described by such a simple formula. 
However, we managed, by fitting appropriate values of constant $b, $to obtain 
neutron stars with quite plausible parameters (see Appendix 1). By varying 
constant $b$ we can, in particular, vary the Oppenheimer---Volkoff limit for 
cool nonrotating neutron stars. We put $M_{OV }= 2 M_{\odot}$ in this paper 
for cool nonrotating neutron stars.

Of course, one must bear in mind that the real Oppeheimer---Volkoff limit 
depends both on the velocity of rotation of the neutron star and on its 
thermal energy (Friedmann et al., 1985 ). In our model this dependence is 
qualitatively consistent with the numerical results obtained earlier. 

We finally introduce dissipative force $F_{diss}$:

\begin{equation}
F_{diss} =-\frac{1}{\tau }\left( {\frac{dR}{dt}} \right)
\end{equation}

It is clear from physical viewpoint that after reaching the centrifugal 
barrier the core undergoes extremely strong oscillations with a time scale 
of $1/\omega $. This process is accompanied by the redistribution of angular 
momentum and complex nonaxisymmetric motions, which must ultimately result 
in the release of half of the gravitational energy and formation of a 
quasi-static cylindrically symmetric object --- a spinar. A detailed analysis 
of this transition is beyond the scope of our simple model. We just 
introduce a damping force assuming that its work transforms entirely into 
heat so that our model correctly describes the total energy release during 
the formation of the spinar, but is absolutely unable to describe the 
temporal behavior at that time. We actually assume that:

\begin{equation}
\tau ={2\pi \chi }/{\omega} 
\end{equation}

Throughout this paper, $\chi =0.04$ unless otherwise indicated.

\subsection *{b) Angular momentum loss equation}

The decrease of the angular momentum of the spinar (collapsar) is due to the 
effect of magnetic and viscous forces. In this paper we assume that 
dissipation of angular momentum is due to the effective magnetic field. In 
this case, the breaking torque in a disk-like object is equal to (see 
Lipunov, 1992)

\begin{equation}
K=\int\limits_{R_{\min } }^\infty \frac{B_z B_\phi\, dS}{4\pi 
}=\frac{1}{2}\int\limits_{R_{\min}}^\infty {B_z } B_\phi R\,dR,
\end{equation}

where $B_z$ and $B_\phi$ --- $z$ and $\phi$ are the components of 
magnetic field.

We now introduce the magnetic moment $\mu$ of the spinar. Hereafter, for 
the sake of simplicity, we write our equations as if the spinar had a dipole 
magnetic field. However, our equations remain unchanged if we simply use 
some average magnetic field of the spinar and characterize this field by the 
spinar magnetic energy $U_{m}$ mentioned above. This is true for the breaking 
torque that we use below.

Let $B_z B_\phi =\kappa_t \, B_d$, where $B_d =\frac{\mu}{R^3}$ is dipolar strength 
of the magnetic fields. The breaking torque is then equal to (see Lipunov, 
1987, 1992 see below)

\begin{equation}
K=\kappa_t\, \frac{\mu^2}{R_t^3},
\end{equation}

where $\kappa _t \sim 1$ and $R_{t}$ is the characteristic radius of 
interaction between the magnetic field and ambient plasma:

\begin{equation}
\begin{array}{l}
R_t = R_{Alfven} \, \textnormal {    is the Alfven radius (Propeller)}\\
R_t = Rc=\left({\frac {GM}{\omega ^2}}\right)^{1/3} \, \textnormal{is the corotation radius (Accretor)}\\
R_t = R_l =\frac{c}{\omega} \, \textnormal { is the radius of the light cylinder (Ejector)}
\end{array}
\end{equation}

In the case of a spinar the Alfven radius is smaller than or on the order of 
the stellar radius and is of little importance in the situation considered.

In the case of a collapsing core the effective interaction radius must be 
close to the corotation radius, which, in turn, is close to the spinar 
radius in accordance with tits equilibrium condition. Therefore the 
retarding torque can be written as:

\begin{equation}
K=\frac{\kappa_t \mu ^2\omega ^2}{GM}=\frac{\kappa_t \mu ^2}{R_B^3 }
\end {equation}

And the corresponding dissipation time scale is:

\begin{equation}
T_C =\frac{I_B \omega ^2}{\mu ^2/R_B^3 }
\end {equation}

Hence the equation of variation of the spinar angular momentum becomes 
(Lipunov, 1987): 

\begin{equation}
\frac{dI\omega }{dt}=-\frac{\mu ^2}{R_c^3 }=-\frac{\kappa _t \mu ^2\omega 
^2}{GM}
\end {equation}

Some authors (Woosly, 1993; Narayan et al., 2001) consider a situation where 
accretion continues onto the newborn black hole at a rate of up to 
$10^{-1}$M/yr. In just the same way accretion may continue onto the spinar. 
The equation of the variation of the angular momentum of the an accreting 
spinar was first derived by (Lipunov, 1987 equation 123):
$$
\frac{dI\omega }{dt}=-\frac{\mu ^2}{R_c^3 }=-\frac{\kappa _t \mu ^2\omega 
^2}{GM}+\mathop M\limits^\bullet \sqrt {GMR} 
$$
where $\mathop M\limits^\bullet$ is the disk-accretion rate. It was shown 
in the same paper that accretion dose not change dramatically spinar 
evolution and hereafter we neglect accretion. The effect of accretion should 
always be important if the accretion time is much shorter than the time 
scale of dissipation of angular momentum, $t_{acctretion} \sim \frac {M_{core}} 
{\mathop {M}\limits^\bullet}<T_C$ However, in this case the very process 
of accretion is the process of the formation of the spinar. Note that our 
scenario differs substantially from that of Woosley (1993), who considers 
accretion to be of importance, because it is the process that determines the 
energetics of the gamma-ray burst. A spinar is a collapsing (but not a 
collapsed!) stellar core. 

In other words, a spinar is by itself an ``accretion disk''. Of course we 
may complicate the model in the future, but we prefer to stop our coarse 
(but physically transparent) approximation here and ignore accretion.

The retarding torque written in this form gives the absolute upper limit for 
the possible spin-down of the spinar.

If the mass of the spinar is below the Oppenheimer---Volkoff limit, a 
neutron star forms ultimately, which spins down in accordance with the 
following magnetodipole formula:

\begin{equation}
\frac{dI\omega}{dt}=-\frac{\kappa_t \, \mu^2}{R_l^3}
\end{equation}

\subsection *{c). Magnetic Field Evolution}

As is well known (Ginsburg and Ozernoy, 1963) magnetic field must disappear 
in the process of collapse.

In the Newtonian approximation in the case of magnetic-flux conservation, 
the dipole moment behaves as:

\begin{equation}
\mu \sim BR^3\sim BR^2R\sim R
\end{equation}

With relativistic effects taken into account, magnetic field vanishes not at 
zero, but when the star reaches the event horizon. Manko and Sibgatullin 
(1992) computed the evolution of the dipole magnetic field of a rotating 
body (in the Kerr metrics). 

We can use the following simple formulas as the first approximation:

\begin {equation}
\mu =\mu_0 \frac{ R-R_{\min}/2 }{R_0-R_{\min}/2} 
\end {equation}

Here $R_{\min}$ is the equatorial radius of the event horizon. Given that 
$R_{0} \gg R_{min}$ , this formula correctly describes the behavior of the dipole 
moment and yields zero magnetic field at the event horizon.

However, this law implies too fast decrease of magnetic field and we use the 
following modified law of magnetic-field decay adopted from Ginsburg and 
Ozernoy (1963):

\begin {equation}
\mu \sim \mu_0 \left( {\frac{R_0}{R}} \right)^2\frac{\xi (x_0 )}{\xi (x)}
\end {equation}

where $\xi (x)=\frac{x_{\min } }{x}+\frac{x^2_{\min } }{2x^2}+\ln \left(1-\frac{x_{\min 
}}{x}\right)$ and $x_{\min}$ is the radius of horizon for current Kerr parameter.

In this paper we neglect the effects of generation of magnetic fields.

\subsection * {d).Energy losses}.

The release of energy in the process of collapse is initially due to the 
dissipation of kinetic energy of the impact onto the centrifugal barrier and 
to spinar spin-down due to magnetic forces:

\begin {equation}
L_0 =\frac{1}{\tau }M\,\left( {\frac{dR}{dt}}\right)^2 \textnormal{  before the formation 
of the spinar}
\end {equation}

\begin {equation}
L_0 =\frac{\mu ^2}{R_{\min }^3 }\omega \textnormal{    after the formation of the spinar }
\end {equation}

Where invariably $R_{\min}=R_{c}$ if the core mass exceeds the 
Oppenheimer---Volkoff limit.

A distant observer would record lower luminosity because of gravitational 
redshift and time dilatation. 

We adopt the following observed luminosity:

\begin {equation}
L_\infty =\alpha ^2L_0
\end {equation}

where $\alpha$ is the time dilatation function --- the ratio of the clock 
rate of reference observers to the world time rate at the equator of the 
Kerr metrics (Thorne et al., 1986):

\begin {equation}
\alpha =\sqrt {\frac{x^2+a^2-2x}{x^2+a^2}}
\end {equation}

If the core mass is below the Oppenheimer---Volkoff limit, the spinar 
ultimately evolves into a neutron star and its luminosity is given by the 
following magnetodipole formula:

\begin {equation}
L_0 =\kappa _t \frac{\mu ^2}{R_l^3 }\omega
\end {equation}

We finally consider the case where rotation is so slow that the spinar does 
not form at all.

In this case direct collapse occurs. We pointed out above that Lipunova 
(1997) was the first to address the problem of electromagnetic burst with 
the allowance for general relativity effects. In the case of direct collapse 
rotational motion is of no importance, because the star makes less than a 
single rotation before it is under the event horizon. 

\begin{figure*}
\hbox to \hsize{
\psfig{figure=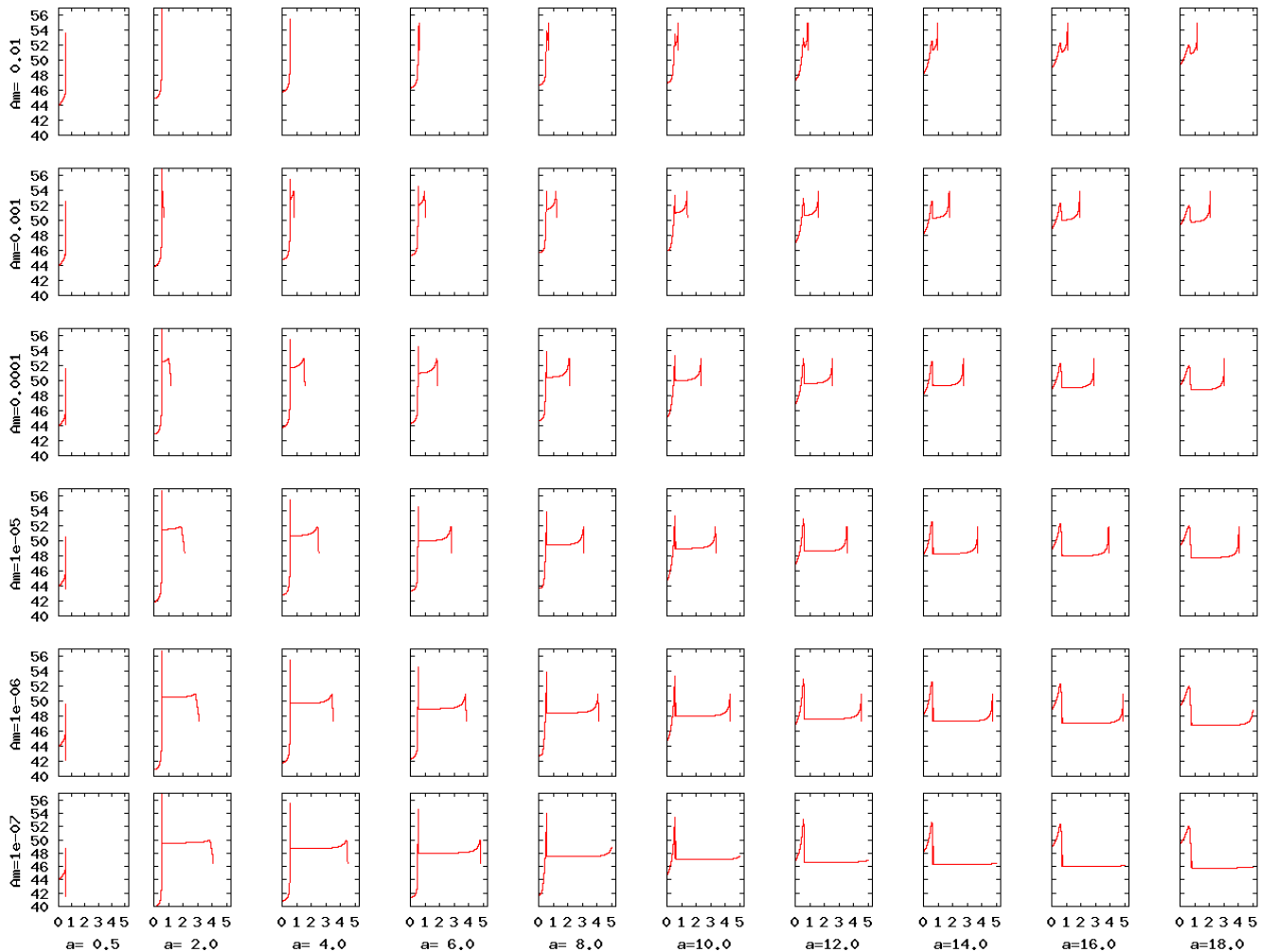,width=180mm}}
\caption{
 Operation of the central engine.
Results of the computation of energy release (luminosity-time in logarithmic
coordinates) during the collapse of 7 $M_\odot$ stars.
}
\end{figure*}


However, this case is 
characterized by large radial variation of the dipole moment:

\begin {equation}
L=\frac{2}{3c^2}\left({\frac{d\mu }{dt}}\right)^2
\end {equation}

To convert this value into the observed luminosity we must take into account 
gravitational redshift and the Dopple effect due to the emitter falling in a 
virtually Schwartzshieldian metrics (Lipunova, 1997).

\section {Collapse of a massive core ($M>M_{OV}$).}

Let us first consider the case where the core mass exceeds substantially the 
Oppenheimer---Volkoff limit. We adopt the initial core mass of $1000Rg$ as the 
initial conditions for our set of differential equations. Figure 3 shows the 
computed variation of the radius, Kerr parameter, and average magnetic field 
for several arbitrary initial core parameters as functions of proper time 
(without the allowance for the time dilatation factor). Diagram 4 shows the 
computed evolution of the central engine for a wide range of models. Let us 
emphasize several important points. First, the collapse of such cores ends 
by the formation of an extremely rotating Kerr black hole. Of course, this 
event shifts to infinitely distant time in the rest frame. 

The diagram (Fig.4) fully corroborates our qualitative scenario (Fig. 2) and 
demonstrates a large variety of the time scales and energies of precursors, 
gamma-ray bursts, and flares. The results of computations of the energy 
release in direct collapse $(a_{0}<1)$ confirms the short duration and low 
power of the flare. Note that the total energy does not exceed 
$10^{-4}Mc^{2}$ for almost all values of magnetic field. Evidently, in this 
case the appearance of jets and of the gamma-ray burst phenomenon is 
difficult to imagine. Such a collapse would rather result in a common 
supernova event.

However, the events acquire an increasingly dramatic turn with increasing 
moment. At $a_{0}>1$ centrifugal forces sooner or later exceed the 
gravitational forces, halt the collapse to give time and opportunity for 
enormous energy of about $\sim 0.1Mc^{2}$ to be radiated during the halt of the 
collapse. In this case a spinar is born and the relativistic jet penetrates 
the envelope of the star and triggers a gamma-ray burst. The magnitude of 
the first burst depends on the initial spinar radius exclusively, which to a 
first approximation is determined only by the moment, as is evident from the 
diagram. All systems in the same column have the same burst energy. Magnetic 
field then takes the reigns of government and determines the rate of 
dissipation of angular momentum and becomes the main factor to determine 
further evolution of the core. As magnetic moment decays, the core radius 
decreases and magnetic luminosity increases until it reaches its maximum (at 
$R\sim Rg$) whose magnitude is determined by the magnetic field exclusively. This is 
also evident from the diagram. After that the luminosity decreases abruptly 
because of relativistic effects near the event horizon (decay of the field, 
gravitational redshift, and time dilatation). It is the ratio of the 
energies of the first and second flare that determines the entire zoo (all 
the variety) of flare, precursor, and burst events. In the extreme case of a 
strong magnetic field (Eq.23) and comparatively small momentum 
$(1<a_{0}<6)$ both flares are very short, narrow, and separated by a short time 
interval of $\sim $1-10s. It is thus impossible in this case to separate the burst from 
the precursor or flare and we must view the event as a double gamma-ray 
burst.

If we move rightward on the diagram in the direction of increasing momentum 
the initial spinar radius increases (for systems with $\sim 100Rg$ large 
precursors), the gravitational energy released decreases, and the first 
flare becomes weaker. We thus fall into the domain of precursors: 
($\alpha_m \sim 10^{-2}-10^{-4}, 10<a_{0}<20$).
 The greater is the angular momentum 
and the stronger the magnetic field, the greater is the separation between 
the precursor and the gamma-ray burst. Luminosity remains virtually constant 
between the precursor and the gamma-ray burst.

If, on the other hand, we move downward from the domain of double gamma-ray 
bursts, thereby decreasing the magnetic-field strength, increasing the time 
interval between the primary and secondary bursts, and decreasing the 
intensity of the second burst, we come into the extended domain of gamma-ray 
bursts ($\alpha_m \sim 10^{-4}-10^{-7}$, $2<a_{0}<14$). If its initial 
angular momentum is comparatively small, the spinar has an initial radius of 
$\sim 10Rg$ and the first burst must be very powerful. Magnetic field, however, is weak 
and the power of the second burst would suffice only to produce X-ray 
flares. In the case of too weak fields ($\alpha _m \le 10^{-7}$) the 
second burst is virtually absent, allowing some bursts (e.g., GRB070110 and 
GRB050904, see below for details) to exhibit a long ($\sim 10^{4}$) x-ray 
plateau.

Finally, the bottom-right corner is occupied by the systems where the energy 
of neither the first nor the second flare is too low for a gamma-ray burst. 
These cores ($\alpha _m \le 10^{-6}$, $a_{0}>14$) may produce either an x-ray 
burst with a precursor or unusual supernovas.

\begin{figure}[!h]
\hbox to \hsize{
\psfig{figure=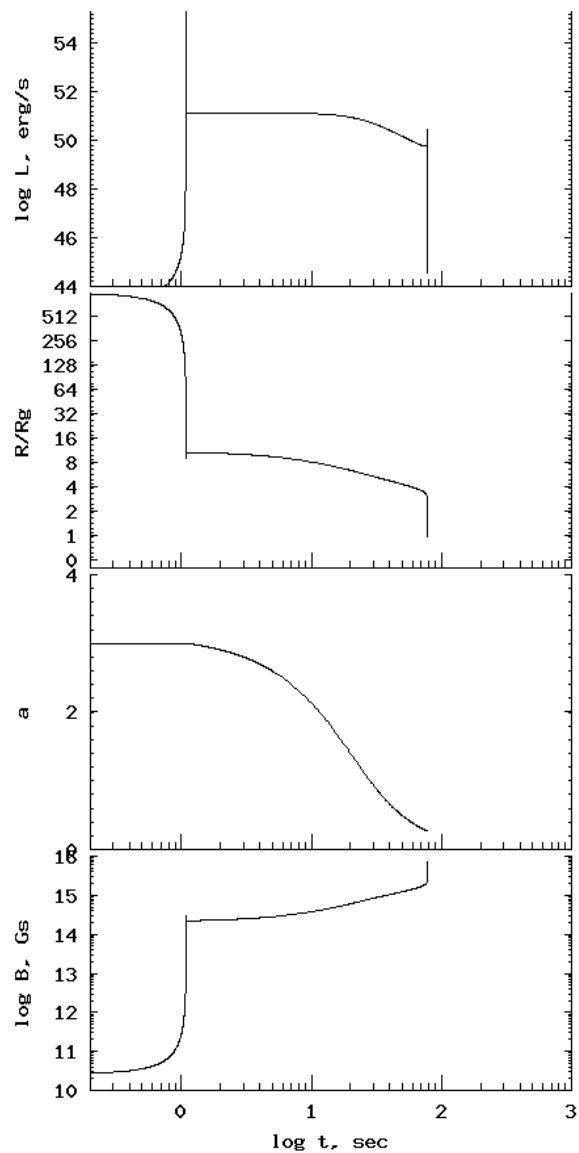,width=80mm,height=170mm}}
\caption{
 Computation of the collapse of a $2.2 M_\odot$
star. The initial effective Kerr parameter and initial magnetic-to-gravitational
energy ratio are equal to $a_0=3$ and $\alpha _m=10^{-4}$ respectively.
}
\end{figure}
\begin{figure}[ht]
\psfig{figure=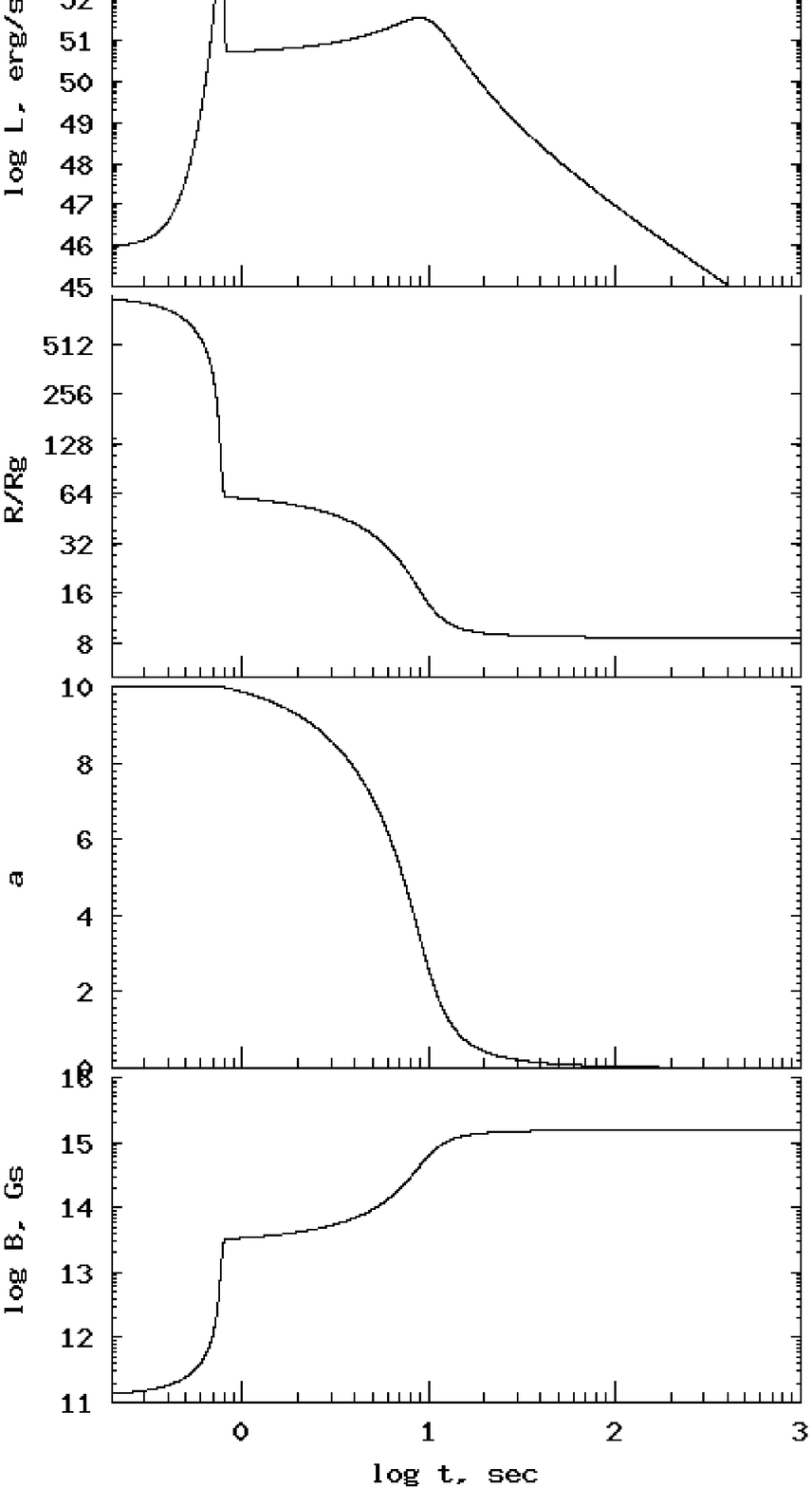,width=80mm,height=170mm}
\caption{
 Computation of the collapse of a $1.5 M_\odot$
star. The initial effective Kerr parameter and initial magnetic-to-gravitational
energy ratio are equal to $a_0 =10$ and $\alpha _m=10^{-3}$, respectively.
}
\end{figure}

\section {Collapse of a rapidly rotating intermediate-mass core 
($M >\sim M_{OV}$)
(supernova case).}

As it was marked (Lipunova, 1997; Lipunova {\&} Lipunov, 1998; Vietri {\&} 
Stella, (1998) - a ``supranova'' scenario), as Oppenheimer---Volkoff limit 
for fast rotating neutron star is higher, then massive NS temporal formation 
is possible. Having lost its rotational moment, the star collapse in to the 
black hole.

For example let us consider a fast rotating core collapse with mass 
$2.2M_\odot $. We should remind that, for distinctness, we use the state 
equation with Oppenheimer---Volkoff limit equal to 2.0 Solar masses (for 
non-rotational neutron star). 

Practically, that means that, spinar equilibrium at last stages of evolution 
is maintained both centrifugal and nuclear forces and may be by thermal 
pressure. 

Fig.5 demonstrates the result of that core collapse calculation. The heavy 
neutron star exists for about 100 seconds. As magnetic momentum looses lead 
to rotation acceleration, its magneto-rotating luminosity after initial 
plateau begins to decrease (the nuclear pressure doesn't allow neutron 
stars-spinar to compress strongly ). 

But after near 100 seconds effective Kerr parameter becomes less than unity 
and relativistic effects result in rapid direct collapse of neutron star 
into the black hole.

\section {Collapse of a rapidly rotating low-mass core 
($M<M_{OV}$).}

The collapse of a low-mass star is ultimately halted by the pressure of 
degenerate matter. However, even in this case fast rotation plays important 
part. In a number of cases, a neutron star does not form directly, but first 
a spinar, which then transforms into a neutron star losing angular momentum. 
Such a collapse does not end by abrupt decrease of luminosity (as in the 
cases considered above), but has a long tail: $L\sim t^{-2}$.

In this example we consider the collapse of a $1.5M_\odot$ core into a 
neutron star (Fig.6). The process results in the formation of a neutron star 
of radius $\sim 8.5Rg (38km)$. Thus the problem acquires yet another characteristic radius 
--- $R_{NS}$ (the nonrotating neutron star radius).

If centrifugal forces less then nuclear pressure 
($R_{NS}>R_{Spinar}$), and this is quite possible with strong fields and small 
angular momenta, the neutron star forms directly and the light curve has 
only one maximum followed by a $t^{-2}$ decrease due to uniform dissipation 
of the angular velocity of the NS. Such systems are located in the 
bottom-left corner of diagram (Fig.7). 
$(\alpha _m \ge 10^{-3},a_{0}<6)$.

If $R_{NS}<R_{Spinar}$, the process again acquires a two-burst pattern. 
However, it does not resemble the collapse of a massive core. This is due to 
the fact that if the radius of the NS is $\sim 10Rg$ no second burst is to be 
expected near $Rg$. Hence we have no systems with precursors and gamma-ray 
bursts occur only in systems with small initial momenta ($a<12$).

\begin{figure*}

\psfig{figure=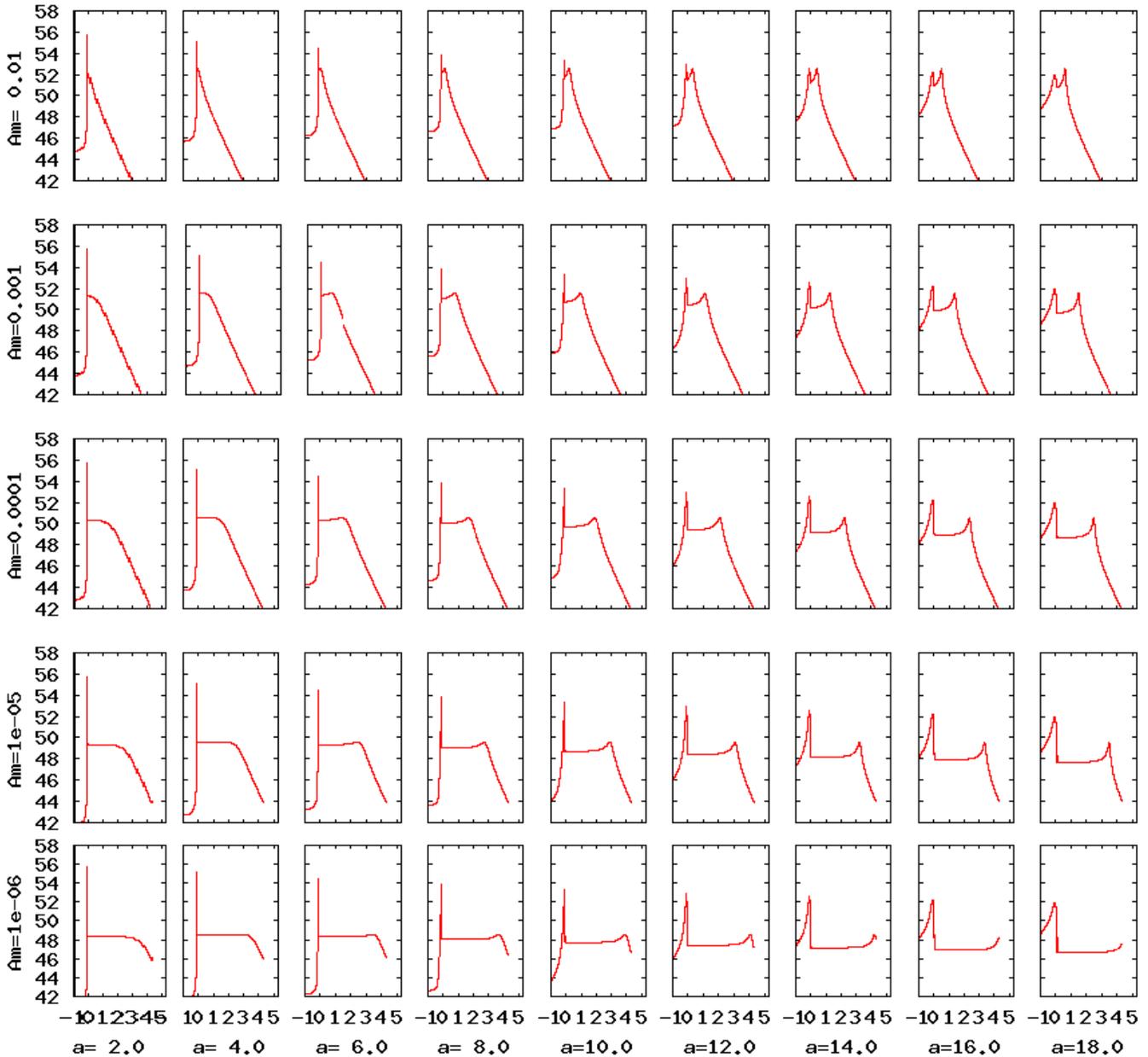,width=180mm}
\caption{
 Operation of the central engine. Results of the computation
of energy release (luminosity-time in logarithmic coordinates) in the
process of the collapse of a core into a $1.5 M_\odot$ neutron
star with different values of the effective Kerr parameter (a) and initial
magnetic-to-gravitational energy ratio ($\alpha_m$). We set the
initial core radius equal to $1000Rg$. The first and second flares correspond
to the formation of the spinar and neutron star, respectively. At the end of
the process, energy release always begins to obey the magnetodipole law
corresponding to the spin-down of the neutron star, i.e., the
pulsar. Computation of the collapse of a 
$1.5 M_\odot$ star. The initial effective Kerr parameter and initial
magnetic-to-gravitational energy ratio are equal to $a_0 =10$ and
$\alpha _m =10^{-3}$, respectively. 
}
\end{figure*}

Some systems with intermediate rotation and strong field 
($\alpha _m \ge 10^{-4},6<a_{0}<12$) 
may produce a weak x-ray flare. This flare is not observed in 
systems with small angular momentum, because in these cases the height of 
the plateau exceeds that of the flare.

In other cases the energy of any flare is hardly sufficient for it to 
penetrate the envelope, and supernovas are observed.

\section {Statistical properties of precursors, flares, and gamma-ray 
bursts.}

An analysis of BATSE data (Lazzati, D., 2005) shows that up to $20$ percent of 
long gamma-ray bursts have precursors preceding the trigger time by up to 
$200 s$. Chincarini et al., (2007) found about 30 flux increase events 
(optical flares) from Swift observatory data.

There are no more doubts that at least a substantial part of these phenomena 
are associated with the peculiarities of the operation of the ``central 
engine''.

\begin{figure}[ht]
\psfig{figure=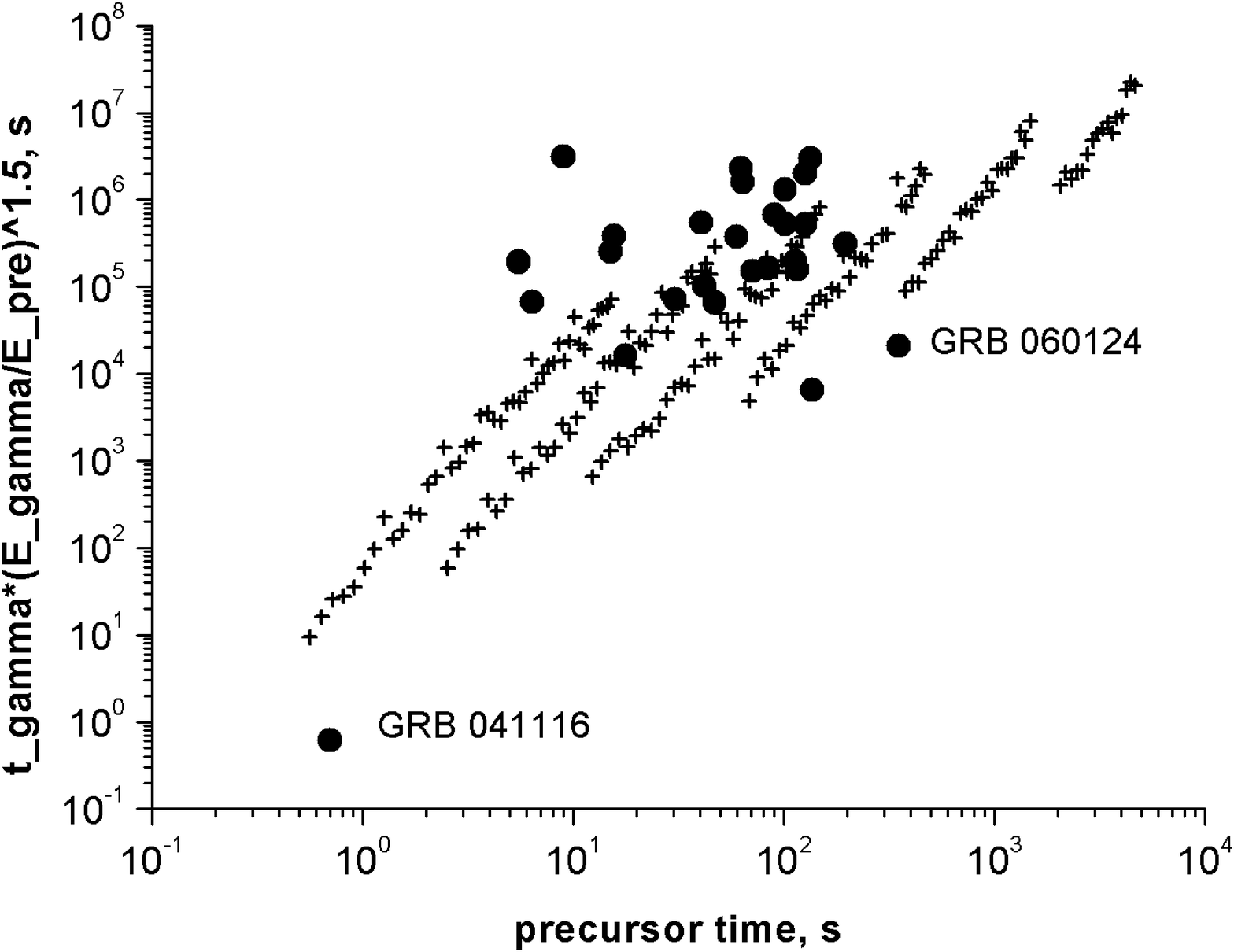,width=80mm}
\caption{
 The gamma-ray burst time multiplied by the gamma-ray-burst to
precursor energy ratio as a function of precursor time. The filled circles
show BATSE data (Lazzati, 2002) and the data for two outsending bursts: a
short (GRB041116) and a long (GRB 060124) one. We use the fluence data and
assume that the opening angles of the precursors are equal to those of the
corresponding gamma-ray bursts. The crosses show the simulated gamma-ray
bursts with precursors computed for a $7 M_\odot$ core. The
effective Kerr parameter varied from 7 to 20 , and magnetic field,
from $10^{-2} to 10^{-6}$.
}
\end{figure}

\subsection {Precursors.}
Numerous observations of gamma-ray bursts show a complex structure in their 
temporal behavior, which is impossible to explain in terms of a single 
burst, formation of a jet, and development of a system of shocks in this 
jet. For example, the model associated with the emergence of the tip of the 
bow shock onto the star's surface (Ramirez-Ruiz et al., 2002; Waxman {\&} 
Meszaros (2003)) can explain precursors that are close to the time of the 
gamma-ray burst (GRB-time), but not the early precursors preceding the main 
GRB by $100-200 s$ (Xiang-Yu Wang {\&} Meszaros, 2007). The latter authors 
proposed a model where early precursors appear as a result of the fallback 
of a part of the star's shell. 

Our proposed scenario naturally explains the phenomenon of precursors and 
flares. In the case of large angular momentum ($a\gg1$) the initial radius 
is large and, correspondingly, the energy release is low, allowing stage 
\textbf{B }to be interpreted as a precursor phenomenon.

In this case, the following condition must evidently be satisfied:
\begin {equation}
T_{pre} =T_B =T_{GRB} \left(\frac{R_B}{Rg}\right)^{3/2}
=T_{GRB}\frac{E_{GRB}}{E_B}
=T_{GRB}\frac{T_{90GRB}}{T_{90Pre}}\frac{\theta _{GRB}^2}{\theta _{Pre}^2}
\end {equation}
where $T_{pre}, T_{GRB}, F_{90}$, and $\theta $ are the observed fluence, and the jet 
opening angle of the gamma-ray burst or precursor, respectively. Note that 
the slope of the latter relation does not depend on the redshift of the 
gamma-ray burst. 

In the model considered a precursor is defined as the initial energy release 
when centrifugal forces halt the collapse of the core (stage B) in the case 
where 
\begin {equation}
{GM}^{2}/R \ll Mc^{2}
\end {equation}
The statistic properties of modeling precursors are presented on Fig.8.

We see more or less good similarity between the artificial and observed 
precursors.

\begin{figure*}
\hbox to \hsize{
\psfig{figure=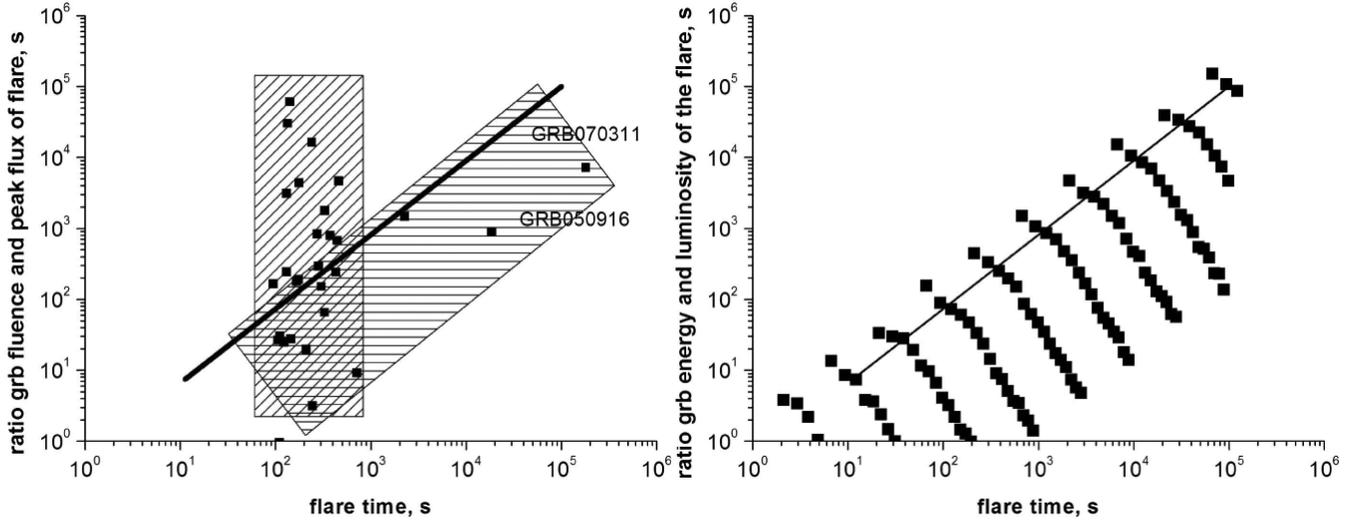,width=180mm}}
\caption{
 The observed GRB fluence-to-peak-luminosity ratio as a
function of flare time (Fig. 9a) based on the data of Lazzati (2005)
supplemented with two interesting bursts GRB 060124 (Romano, P et al 2006)
and GRB 041116 (Golenetskii, et al GCN2835). Theoretical ratio for simulated
gamma-ray bursts (Fig.9b) . In our computations we assume that the core mass
is equal to $7 M_\odot$ the effective Kerr parameter varies from 2 to 7,
and magnetic energy lies between 0.01 to $10^{-7}$.
The solid line is based on bursts, it corresponds rather accurately
to equation $\frac{Fluence_{GRB}} {F_{flare}}=t_{flare}$.
}
\end{figure*}

\subsection {X-Ray Flares.}

If the initial spinar radius R$_{B }$is small, the energy release at the 
time of its formation is sufficient to produce a gamma-ray burst, and hence 
the first flare should be interpreted as a gamma-ray burst, whereas the 
secondary release of energy by the spinar can be interpreted as a flare. In 
this case the energy of the gamma-ray burst is approximately equal to:
\begin {equation}
E_{GRB}=E_B \approx \frac{GM^2}{2R_B} = \left(\frac{1}{2a_0^2}\right)M_{core} c^2
\end {equation}
The burst luminosity is 
\begin {equation}
L_{flare} =\frac{\mu_g^2}{R_g^3}\omega_g 
\end {equation}

where $\mu_{g}$ and $\omega_{g}$ are the magnetic moment and angular 
velocity at the distance of $R_{g}$, respectively:
$$
\mu _g \sim \mu _0 \frac{Rg}{R_0}\\
\omega _g \sim \omega _0 \left(\frac{R_0}{Rg}\right)^2
$$
Hence the time gap between the gamma-ray burst and the flare is equal to the 
spin-down time of the spinar at the maximum radius:
$$
t_{flare} \sim I_B \omega R_B^3 /\mu _B^2
$$
Simple substitutions yield the following relation between the observed flare 
parameters:
\begin {equation}
L_{flare} =\frac{E_{GRB}} {t_{flare}}(\frac{R_B}{Rg})^{5/2} \sim \frac{E_{GRB}} {t_{flare}}\left(\frac{Mc^2}{E_{GRB}}\right)^{5/2}
\end {equation}

We now substitute the observed quantities into the latter formula to obtain: 

\begin {equation}
\frac{Fluence_{GRB}}{F_{flare}}=
\left(\frac{\theta_{flare}^2}{\theta_{GRB}^2}\right)
\left(\frac{Mc^2}{E_{GRB} }\right)^{-5/2}t_{flare}
\end {equation}

where$F_{flare}$ is the maximum flux during the flare and 
$Fluence_{GRB}$ is the total fluence of the gamma-ray burst.

In the latter relation $E_{GRB}$ is the only quantity that depends on the 
distance to the gamma-ray burst. We can therefore plot the observed 
relation $Fluence_{GRB}/F_{flare}=function(t_{flare})$. 
Figure 9 shows the observed 
relation and the relation simulated in our model. We adopt experimental data 
from Chincarini et al. (2007). The straight line shows approximate 
analytical relation (53) :

Figure 9b shows our computed models for the collapse of 7-solar mass star 
with parameters:

\begin{center}
$2<a<20$

$10^{-7}<\alpha _m <10^{-2}$
\end{center}

One can see that the observed and theoretical spectra show similar trends 
for the part of the flares (inclined rectangle): they both grow toward 
(temporally) distant flares with a comparable scatter. The scatter is mostly 
due to the large factor, and the difference between the mean values is due 
to the following ratio 

\begin {equation}
\left(\frac{\theta_{flare}^2}{\theta_{GRB}^2}\right)
\left(\frac{Mc^2}{E_{GRB} }\right)^{-5/2}\sim 10
\end {equation}

When converting XRT observations we assumed that \textit{1 count/s=10}$^{-10}$ (Sakamoto et al. 
2006, GCN Report 19.1 02Dec06)

One must bear in mind, when comparing the observed and simulated points, 
that BAT and XRT soft-ray detectors operate in different energy intervals. 
XRT observations are made in the energy interval $0.3-10 keV$,
where absorption may be 
important. In addition, the observed fluxes during flares must be multiplied 
by a factor of five to seven, because the spectrum has a power-law form and 
is much wider than the XRT energy channel. Moreover, part of the flares 
(especially those with delays $<100 s$) can also be explained by the 
emission of a system of shocks (reverse shock Chincarini et al., (2007)).

All this leads us to conclude that the slope and scatter of the average 
theoretical and observational curves agree well with each other (for part of 
the flares) and the absolute vertical shift may be due to the differences of 
the directivity diagrams of the gamma-ray burst and optical flare, soft 
x-ray extinction, and extrapolation of the power-law spectrum to a wider 
energy interval.

\begin{figure}[ht]
\psfig{figure=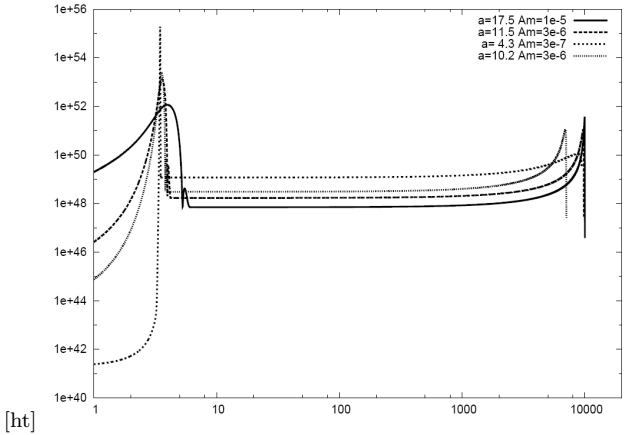,width=80mm}
\caption{
 Computed energy release during the process of collapse with
low angular momentum and weak magnetic field.
}
\end{figure}

\begin{figure}[ht]
\psfig{figure=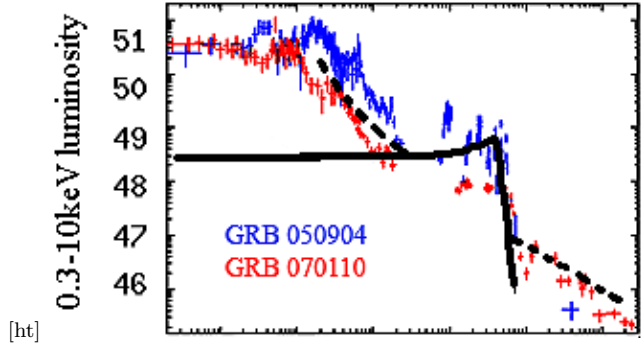,width=80mm}
\caption{
 Experimental Swift X-Ray light curve of the long curve on
the x-ray plateau (Troja et al. (2007))and theoretical model luminosity with
parameters: $Umag/Ug = 10^{-7}$ and effective Kerr
parameter $a=2.0$ (black line).
}
\end{figure}

\section {An extraordinary long X-ray plateau GRB070110 and GRB050904.}

Two of several hundred gamma-ray bursts --- 
$GRB070110$ and $GRB050904$ 
--- do not fit the common scenario of the X-Ray afterglow formation. Both 
bursts exhibit a long plateau with a rest-frame duration of $6000-7000 s$. 
Troja et al. (2007) associated such a long activity with the specifics of 
the central engine and, in particular, with the formation of a neutron star 
after the collapse of a low-mass core (with the mass below the 
Oppenheimer---Volkoff limit). 

We fully agree that such an unusual behavior of the X-ray afterglow is due 
to the central engine, but we believe that the plateau appears not as a 
result of the radiation of the neutron star, but as a result of the activity 
of a spinar with anomalously weak magnetic field. An hypothesis Troja, E., 
et al. (2007) associates the plateau with the collapse producing a neutron 
star -- a radio pulsar -- whose activity becomes appreciable during the 
fading of the afterglow. We believe this interpretation of the plateau to be 
too far fetched. The intensity of the magnetodipole radiation, which is 
typical for radio pulsars, decreases with time as $t^{-2}$. The abrupt 
termination of the plateau stage remains completely unexplained in the young 
pulsar model. The authors of this hypothesis point out that the decrease of 
luminosity could be a result of generation of the magnetic field. However, 
the last assumption makes the model too complicated.

A plateau with a slight increase and abrupt decrease of luminosity appears 
naturally in the spinar paradigm.

\begin{figure*}
\psfig{figure=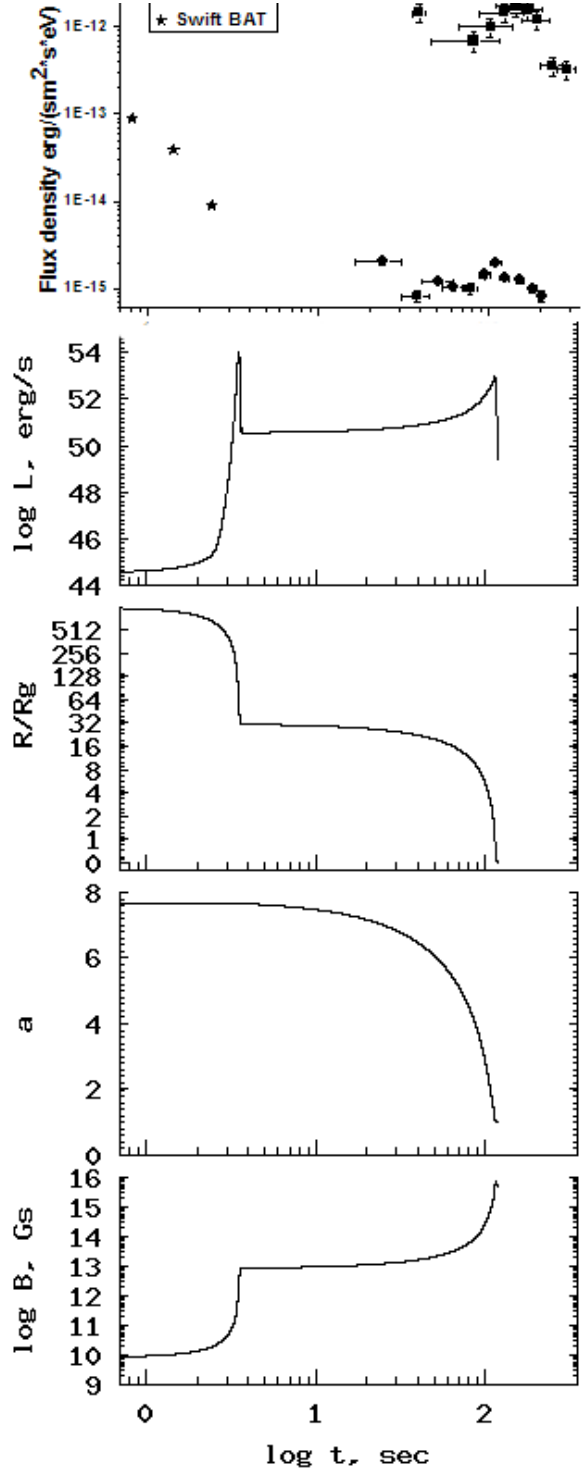,width=80mm}
\caption{
 Light curve and theoretically computed luminosity,radius , effective Kerr
parameter, and magnetic field for GRB060926. The initial parameter
for theoretical computation is
$a_0=7.6$ and $\alpha _m=10^{-4}$.
}
\end{figure*}

In our scenario a plateau is a flare with weak magnetic field. In other 
words, in this case the gamma-ray burst corresponds to the halt of the 
collapse by centrifugal forces at radius R$_{B }$, and the plateau is an 
extended flare due to magneto-rotational losses.

Let us first make some approximate estimates. The initial Kerr parameter is 
equal to:
\begin {equation}
A_0=\frac{I\,\omega\, c}{GM^2}
\end {equation}
The initial spinar radius is:
\begin {equation}
R_{s}=\frac{a^{2}GM}{c^2}=\frac{1}{2}a^2\,Rg
\end {equation}

The energy of the gamma-ray burst is
\begin {equation}
E_{GRB}=\frac{1}{2}\frac{GM^2}{R_{s}}=\frac{Mc^{2}}{2a^{2}}
\end {equation}
We derive from this relation the Kerr parameter:
\begin {equation}
a=E_{GRB}/Mc^{2}
\end {equation}

The characteristic plateau duration is determined by the time scale of the 
loss of the spinar angular momentum:
\begin {equation}
t_{plato} =t_{flare} =\frac{I_B \omega _B }{\kappa _t \textstyle{{\mu ^2} 
\over {R_B^3 }}}\sim \frac{GMa^3}{2\kappa _t c^3\alpha _m }
\end {equation}

The luminosity of the plateau at its maximum computed without the allowance 
for relativistic effects is:
\begin {equation}
L_{plato} (\max )\sim \frac{\alpha _m \kappa _t }{4x}\frac{c^5}{G}
\end {equation}

We now use the observed plateau time to derive the parameters of the 
collapse:
\begin {equation}
\alpha _m =\frac{GMa^3}{2\kappa_t c^3t_{plato} }\sim 10^{-8}\left( 
{\frac{M}{10M_\odot }} \right)t_4 ^{-1}a^3\kappa_t ^{1/3}
\end {equation}

Figure 10 shows the theoretical curve of the luminosity spinar evolution for 
different initial parameter. This spinar light curve shows a characteristic 
plateau whose luminosity and duration are totally consistent with 
experimental data (Troja et al., 2007). So the plateau appears naturally in 
the spinar model and it is a typical feature for the collapse of a core with 
small angular momentum and weak magnetic field, and may be find in many 
different cases.

To illustrate these points, we computed a best theoretical light curve 
($\alpha_{m} = 10^{-7}, a=2.0$) artificially supplying additional self-similar radiation 
in accordance with the following law (Fig.11) and preview if with 
experimental data given by Troja et al. (2007):

$F=F_{theory}+ C_{1} t^{-2}+ C_{2 }t^{-1}$

\begin{figure}[ht]
\psfig{figure=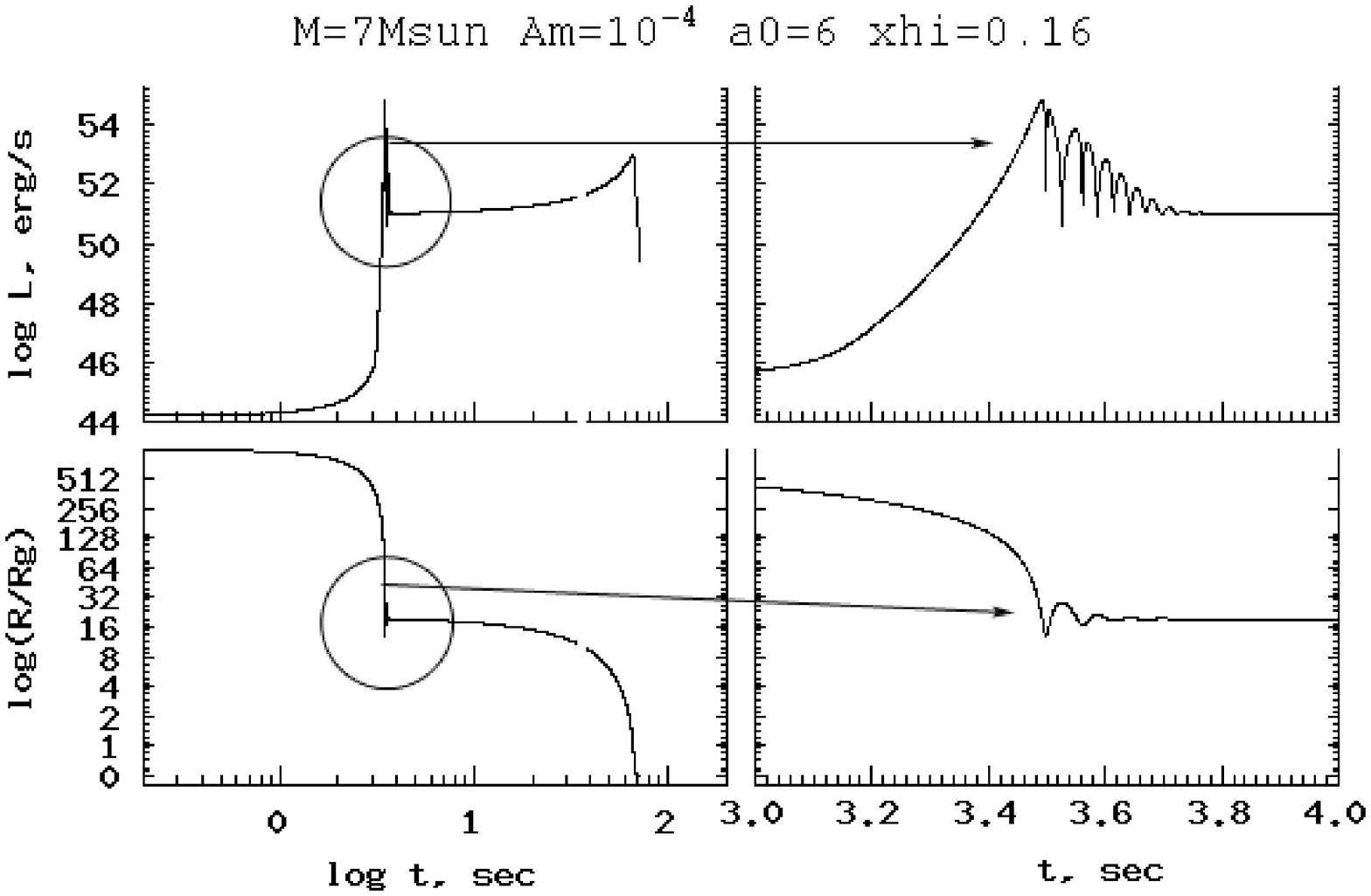,width=80mm}
\caption{
 Computed collapse of a $7 M_\odot$ core.
Qualitative illustration of the fine structure of the temporal behavior of
the gamma-ray burst or multiple precursors.
}
\end{figure}

\section {GRB 060926}

X-ray flares can sometimes also be observed at optical wavelengths. Let us 
try to explain the phenomenon of such a flare in the gamma-ray burst GRB 
060926, where an optical flare was discovered along with the x-ray flare. We 
choose this burst not only because we want to illustrate how spinar paradigm 
works for flares observed both at x-ray and optical wavelengths, but also 
because the optical radiation of this burst was discovered by MASTER group 
whose members include the authors of this paper. 

Optical observations of the gamma-ray burst $GRB060926$ recorded by Swift 
gamma-ray observatory (Holland,S et al 2006) were performed with MASTER 
telescope operating in an automatic mode under good weather conditions 
(Lipunov et al 2006). The first exposure started at 16:49:57 UT 2006-09-26, 
$76 s$ after the gamma-ray burst was recorded. We found in the first and 
subsequent coadded frames an optical transient with the following 
coordinates:

$\alpha =17^{h}35^{m}43^{s}.66$

$\delta =13^{d}02^{m}18^{s}.3$

$err=\pm0^{s}.7''$

which agree with the coordinates of the optical transient discovered by 
Holland et al. (2006) within the errors of our observations. The results of 
the corresponding photometry yielded the first data points on the light 
curve.

We found an optical flare event --- after a short decrease the brightness 
began to rise beginning with the $300^{th}$ second and reached its maximum 
near $500--700 s$. Synchronous X-ray flux measurements with Swift XRT show a 
similar event (see Fig.12). Note that the absorption determined from x-ray 
data corresponds to a column density of $n_{H}=2.2 10^{21} cm^{-2}$ 
of which $n_{H}=7\dot10^{20} cm^{-2}$ 
is Galactic absorption (Holland et al. 2006). Given the 
redshift $z=3.208$, the total absorption in our band is equal to three magnitudes. We 
naturally assume that the dust-to-hydrogen ratio is the same as in our 
Galaxy. A comparison of our optical measurements with the x-ray fluxes 
measured by Swift XRT (Holland et al. 2006) allowed us to determine the 
slope of the spectrum, which we found to be constant within the errors and 
equal to $\beta=1.0 \pm 0.2$: 

\begin {equation}
F \sim E^{-\beta}[erg/cm^{2}s\,eV]
\end {equation}

The spectrum obtained agrees with the x-ray spectrum within the errors 
(Holland,S et al 2006). 

Such a phenomenon was already observed at least in several cases: GRB060218A 
z=0.03 (Quimby et al, 2006a, GCN4782) at the 
$1000^{th}$ second, GRB060729 z=0.54 at the 
 $450^{th}$ second ( Quimby et al., 2006b,c GCN 5366,5377), 
GRB060526 z=3.21 at the $188^{th}$ second (Dai X. et al 2007), and also during the 
bursts GRB990123, GRB041219a, GRB060111b (Wei D.M., 2007) , etc. 

Note that the gamma-ray burst that we discuss here has a redshift of $3.208$ 
(V.D'Elia et al GCN5637). Figure 12b-e shows the results of optical and 
X-ray observations of the flare and of the theoretical computations of a 
spinar with parameters $a_{0}=7.6$ and $\alpha_m =10^{-4}$.

Note also that redshift dilates all time intervals by a factor of $(1+z)$ and 
therefore we show all experimental light curves reduced to the rest-frame. 
We hence have to explain a flare at the $\sim 100^{th}$ second, which is 
about 50 times weaker than the gamma-ray burst as we illustrate in Fig.12.

\begin{figure}[ht]
\psfig{figure=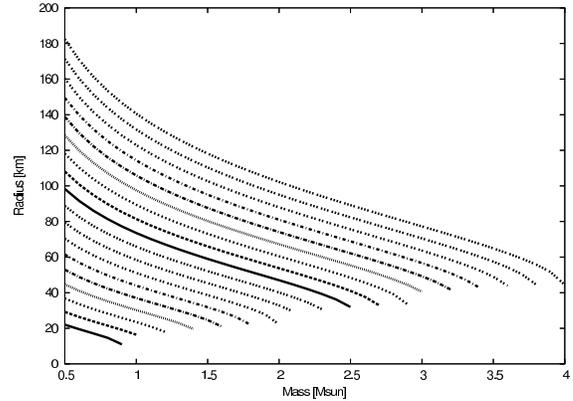,angle=-90,width=80mm}

\label{fig14}
\caption{
 Show the radius of a nonrotating neutron star depends
on its mass for various values of parameter $b$
that appears in our equation of state.
}
\end{figure}

\begin{figure}[ht]
\psfig{figure=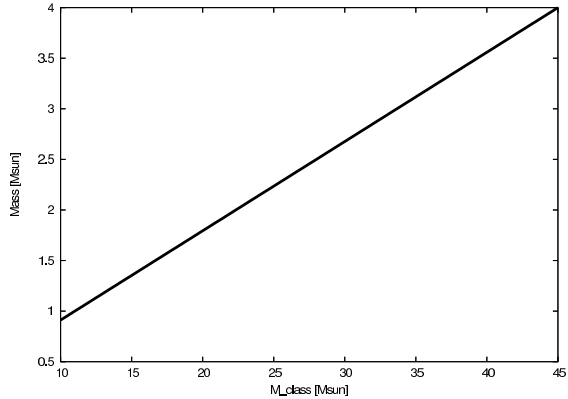,angle=-90,width=80mm}

\caption{
 The Oppenheimer---Volkoff limit as a function of parameter
$M_{Class}\,\, see (eq.29)$ that appears in our equation of state (28) for a
nonrotating neutron star.
}
\end{figure}

\section {Discussion}

Our proposed non-stationary model developed in terms of the Spinar Pardigm 
of magneto-rotational collapse is physically transparent. It takes into 
account all the main relativistic effects and allows their impact on the 
operation of the central engine and on the accompanying events to be 
estimated. It goes without saying that this model cannot replace precise 
magnetohydrodynamic computations, but it evidently helps to choose the 
inevitable simplifications for such computations.

The central assumption in our model is that dissipation of the angular 
momentum of the collapsing core is due to magnetic field. It is clear that 
turbulent viscosity and generation of Alfven waves may play important part 
in the real situation. However, on the one hand, no simple physical model 
has so far been developed for these events and, on the other hand, the 
magnetic field that we introduce can be viewed as some effective parameter 
describing viscous loss of momentum. We point out that although we use 
dipole moment in our set of equations of motion, they actually do not assume 
the dipole nature of the magnetic field. This remarkable circumstance is due 
to the fact that the spin-down torque $\mu^2/R_c^3$ that we use here 
coincides with the energy of magnetic field, $U_m \approx \mu^2/R_c^3$, 
for a spinar whose radius is equal to the corotation radius $R = R_{c}$. By the 
way, this fact proves that we adopted maximally effective spin-down magnetic 
moment.

To reduce the number of initial hypotheses, we never allowed for the 
possible generation of magnetic field (Kluznuzk {\&} Ruderman, 1998) as a 
result of differential rotation of the collapsing core. On the other hand, 
generation of magnetic field can be easily incorporated into the 
approximation employed. It can be done should theory clearly disagree with 
observations. 

There are other phenomena capable of complicating the picture described 
above. For example, the spinar may at a certain time break into two objects 
during the collapse of a rotating core (Berezinski et al.,1988; Imshennik, 
1992). We do not yet consider the second possibility, which, in principle, 
may result in the appearance of several flares or precursors around the 
gamma-ray burst.

The problem of precursors and flares requires a separate explanation. On the 
one hand, we stress that close precursors and flares may result from the 
presence of a complex system of shocks in the relativistic jet. On the other 
hand, the phenomenon of multiple precursors can be easily explained by the 
oscillations of the newborn spinar at the centrifugal barrier. We 
artificially suppressed these oscillations by introducing a special 
dissipative force with the dissipation time scale parameter. As we showed 
above, we can obtain up to 10 precursors for a single gamma-ray burst if we 
choose the dissipation time scale to be one order of magnitude longer than 
the period of spinar rotation (Fig.13).

However, the description of fine effects lies beyond the scope of this 
paper.

We assume that interpreting shock events accompanying the gamma-ray bursts 
in terms of a simple two-parameter scheme is an important step toward 
understanding the operation of the central engines of the gamma-ray bursts.

\textbf{We are grateful to the Russian Foundation for Basic Research for 
having discontinued the financial support of our experimental studies of 
gamma-ray bursts with the first Russian MASTER robotic telescope and thereby 
giving us time to write this paper. We are also grateful to Pavel Gritsyk 
for discussions and assistance in computations and to an anonymous referee 
for useful comments.}

\begin{figure}[ht]
\psfig{figure=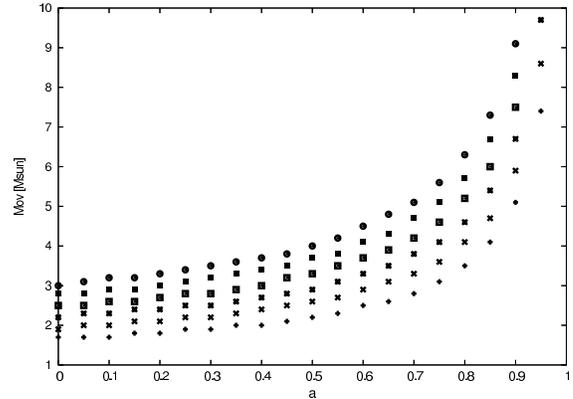,angle=-90,width=80mm}

\caption{
 The Oppenheimer---Volkoff limit as a function of the
velocity of rotation of the neutron star (in the units of the Kerr 
parameter).
}
\end{figure}

\appendix {Appendix 1. Parameters of neutron stars with equation of state 
(28).}

To choose the most appropriate constants in approximate equation of state 
(28), we analyze the global properties of neutron stars in accordance with 
dynamic equation (24), which in the static case transforms into the 
following simple equation:

$\begin{array}{l}
 \frac{4}{Rg^2}\frac{GM}{x^3}\frac{(x^2-2ax+a^2)^2}{(\sqrt x 
(x-2)+a)^2}-\omega ^2R-\frac{P}{R}=0{\begin{array}{*{20}c}
 \hfill & \hfill & \hfill & {=>} \hfill \\
\end{array} } \\ 
 \frac{R}{2MGRg}\frac{(4M^2G^2-2MG\omega c\sqrt {2Rg^2R} +Rg^2R^2\omega 
^2c^2)^2}{(2MG\sqrt {2\frac{R}{Rg}} (R-Rg)+RgR^2\omega c)^2}-\omega 
^2R-\frac{P}{R}=0 \\ 
 \end{array}$

Figure 14 shows the dependence of the radius of a nonrotating neutron star 
on its mass for various values of parameter $b$, which appears in our equation of 
state. First, we see natural decrease of the star's radius with increasing 
mass, which is typical of self-gravitating configurations with equilibrium 
maintained by the pressure of ideal degenerate gas. However, this is of 
minor importance for us compared to the fact that the radii of neutron stars 
of reasonable ($1.5-3M_\odot$) masses lie within reasonable limits: from 
20 to 100 km. Figure 15 shows the dependence of the Oppenheimer---Volkoff 
limit on our parameter $b$. The available orthodox model equations of state for 
neutron stars predict that the maximum mass of a neutron star is 
$1.5-3M_\odot$. This corresponds to the following interval of parameter 
$b$: Fig. 16 demonstrated Oppenheimer---Volkoff --limit depend on effective Kerr 
parameter.

$20M_ \odot<b<35M_ \odot $

Fig. 16 demonstrated Oppenheimer---Volkoff ---limit depend on effective Kerr 
parameter.

\clearpage 

\newpage

\newpage

\clearpage

\clearpage

\newpage

\clearpage

\newpage

\newpage

\end{document}